\documentclass{article}
\usepackage{a4wide,amssymb,amsmath,cite}
\title{Improved Variational Analysis of Deconfinement in $SU(N)$ Gauge Theory}
\author{B. M. Gripaios\thanks{b.gripaios1@physics.ox.ac.uk}\\ 
\emph{Department of Physics - Theoretical Physics, University of Oxford,}\\ 
\emph{1, Keble Road, Oxford. OX1 3NP  UK}\\
J. G. Milhano\thanks{gui@nat.vu.nl}\\
\emph{Department of Physics and Astronomy, Vrije Universiteit}\\
\emph{De Boelelaan 1081, NL-1081 HV Amsterdam, The Netherlands}
\\
\\ OUTP-03 07P }
\date{February 19, 2003}
\begin{document}
\bibliographystyle{h-elsevier2}
\maketitle
\begin{abstract}
A variational analysis of the pure $SU(N)$ gauge theory in 3+1 dimensions
at finite temperature is performed, extending the work of Kogan, Kovner and Milhano \cite{Kogan:2002yr}.
A de-confining phase transition is found at a temperature of 470 MeV, somewhat higher than lattice estimates \cite{Teper:1998kw}. This value is however rather sensitive, for reasons which are discussed. A more robust quantity is the ratio of the transition temperature to the lightest glueball mass in the model. This is $0.18$, in agreement with the lattice estimate for $SU(3)$ to two significant figures. Ways of further improving the calculation are discussed.

PACS Numbers: 12.38.Aw, 12.38.Lg 

Keywords: QCD, Finite Temperature, Variational Approximation
\end{abstract}
\section{Introduction}
In a recent paper \cite{Kogan:2002yr}, a variational method is used to study the deconfinement transition in the pure $SU(N)$ gauge theory 
at finite temperature. The method mimics the Rayleigh-Ritz variational method in the Schr\"{o}dinger formulation of
quantum mechanics. There, the standard procedure is to take a physically motivated ansatz for the ground state wavefunction, 
parameterized by some
free parameters, and to minimise the expectation value of the Hamiltonian with respect to those parameters. 
This provides an upper bound for the ground state (vacuum) energy. The method at finite temperature is analogous: 
in the canonical ensemble formulation of quantum statistical mechanics, one forms an ansatz for the density matrix,
with free parameters, and minimises the expectation value of the Helmholtz free energy. This provides an upper bound
for the free energy at a given temperature.

In section \ref{sec:KKM}, we begin with a pr\'{e}\c{c}is of the approach followed in \cite{Kogan:2002yr}. 
The calculation generalises the variational analysis at zero temperature
performed in \cite{Kogan:1994wf}; an additional kernel $H$ in the ansatz corresponds to the effect of thermal disorder in the system.
This kernel is taken to be small, and only the leading order correction to the entropy of order $H \log H$ is considered.
In this approximation a deconfining phase transition is found to occur at a temperature of 450 MeV.

In section \ref{sec:ext}, we consider higher order corrections in $H$ to the entropy. 
It is shown that (within approximations already present in the Kogan--Kovner model at zero temperature) 
one can calculate the entropy to all orders in $H$ in the high temperature phase. In this extended analysis, the transition temperature is shifted to 470 MeV, which is high compared to lattice estimates \cite{Teper:1998kw}. However, this value depends on a mean-field estimate of the critical coupling in a sigma model which arises in the analysis, and is therefore only approximate. The ratio of the transition temperature to the lightest glueball mass in the model, which is independent of this mean field estimate, is 0.18. This is in agreement with the lattice estimate for $SU(3)$ to two significant figures.

We conclude in section \ref{sec:disc} by discussing our results, and suggesting further improvements.
\section{The order $H \log H$ analysis} \label{sec:KKM}
The ansatz is constructed by considering density matrices which in the field basis have gaussian matrix elements, and where gauge invariance is explicitly imposed by projection onto the gauge-invariant sector of the Hilbert space. It reads
\begin{gather} \label{qcdans}
\rho [A,A^{'}] =  
\int DU \; \exp\bigg\{-\frac{1}{2} \left[ A G^{-1} A + A^{'U} G^{-1} A^{'U} - 2 A H A^{'U} \right] \bigg\}\, ,
\end{gather}
where, under an $SU(N)$ gauge transformation $U$, $A \rightarrow A^U$ and $DU$ is the $SU(N)$ group-invariant measure.
In the above we employ a matrix notation, with e.g.
\begin{gather} \label{matnot}
A G H A = \int dxdydz \; A_{i}^{a} (x) G_{ij}^{ab} (x-y) H_{jk}^{bc} (y-z) A_{k}^{c} (z)\, .
\end{gather}
Here, indices $i,j,k,\dots\ \epsilon\ \{1,2,3\}$ and $a,b,c,\dots\ \epsilon\ \{1,2,\dots,N^2-1\}$ 
denote the spatial Lorentz components and colour components of the gauge field respectively. 
Explicitly, the gauge transformations are
\begin{gather} \label{gt}
A^{a}_{i} (x) \rightarrow A^{U a}_{i} (x) = S^{ab} (x) A^{b}_{i} (x) + \lambda^{a}_{i} (x)\, ,
\end{gather}
with $S^{ab} = \frac{1}{2} \mathrm{tr} ( \tau^a U^{\dag} \tau^b U )$, 
$\lambda^{a}_{i} = \frac{i}{g} \mathrm{tr} ( \tau^a U^{\dag} \partial_i  U )$, 
and 
$\frac{\tau^a}{2}$ form an $N \times N$ Hermitian representation of $SU(N)$: 
$[ \frac{\tau^a}{2} , \frac{\tau^b}{2} ] = i f^{abc} \frac{\tau^c}{2}$ with normalisation 
$\mathrm{tr} ( \tau^a \tau^b ) = 2 \delta^{ab}$.

The kernels $G^{-1}$ and $H$ are arbitrary variational functions. To facilitate the calculation, they are restricted to
be isotropic in colour and space indices. Furthermore, one splits the momenta into high and low modes with $k \lessgtr M$ and restricts
the kernels to the one parameter momentum space forms
\begin{gather} \label{rkers}
G^{-1}(k) = \left\{ \begin{matrix} M, \; k<M \\k, \; k>M \end{matrix}\right. \, ,  \qquad
H (k) = \left\{ \begin{matrix} H, \; k<M \\0, \; k>M \end{matrix}\right.\, .
\end{gather}
The form for $G^{-1}$ is motivated by the propagator for a massive scalar field, \emph{viz.\ }$(k^2 + M^2)^{1/2}$; the form for $H$ assumes
that only the low modes are thermally excited at the temperatures of interest.\footnote{
Non-zero $H$ in (\ref{qcdans}) corresponds to thermal disordering, since $H=0$ corresponds to a pure state.}
With the above restrictions on the kernels,  only two variational parameters, $M$ and $H$, remain.

Before discussing the variational analysis at finite temperature, let us recall the analysis at zero temperature. The former will
turn out to be a straightforward generalisation of the latter. At $T=0$, $H=0$ and the analysis reduces to the minimisation of
the energy, that is of the expectation value of the Hamiltonian, $U = \mathrm{tr} \mathcal{H} \rho / \mathrm{tr} \rho$, where
\begin{gather} \label{ham}
\mathcal{H} = \frac{1}{2} \left[ E^2 + B^2\right]\, ,
\end{gather}
with $E^{a}_{i} = \delta/\delta A^{a}_{i}$ and $B^{a}_{i} = \epsilon_{ijk} (\partial_j A^{a}_{k} + g f^{abc} A^{b}_{j} A^{c}_{k} /2)$.
This is equivalent to the analysis originally performed by Kogan and Kovner in \cite{Kogan:1994wf}.
Firstly, one performs the Gaussian integrals over the gauge fields $A$. 
This leaves integrals over the gauge transformations $U$, evaluated with respect to a sigma model `action'
which is both non-local and non-polynomial in $U$. To simplify the action, the gauge transformations $U$
are split into parts dependent on high and low momentum modes, with $k \lessgtr M$ as above. The effect of integrating out 
the high modes is to effect a renormalisation group transformation: the coupling $g^2$ of the low mode sigma model
is replaced by the renormalised coupling $g^2(M)$. So $M$ acts as a UV cut-off for the low mode theory. 
Furthermore, the theory is asymptotically free \cite{Brown:1997gm,Brown:1997nz}. 
Provided $M$ is sufficiently large (and $g^2(M)$ sufficiently small)
one can then consider the low mode theory to leading order in $g^2(M)$. The relevant Euclidean actions are
\begin{gather} \label{aaction}
S [A] = \left( A + \frac{\lambda}{2} \right) G^{-1} \left(A + \frac{\lambda}{2} \right) + \frac{1}{4} \lambda G^{-1} \lambda\, ,
\end{gather}
for the $A$ fields and
\begin{gather} \label{laction}
\Gamma [U] = \frac{M}{2 g^2(M)} \mathrm{tr_\mathit{{SU(N)}}} \int d^3x \; \partial_i U^{\dag} (x) \partial_i U (x)\, ,
\end{gather}
for the low mode $U$ fields, where the trace is performed over $SU(N)$ matrices $U$. The high modes do contribute to the energy at zero temperature. However, they do not yield any additional contribution at finite temperature,
since $H$, which parameterizes the thermal disorder in the theory, is zero for $k>M$.

Next consider this low mode sigma model as a statistical mechanical system at `temperature' $g^2(M)$. 
The system undergoes a phase transition with spontaneous symmetry breaking from a disordered state
at small $M$ (large $g^2(M)$) to an ordered state at large $M$. Calculations in the disordered phase are performed in the mean field approximation: the $U$ are treated as $N^2$ free fields obeying the unitarity constraint
$U^{\dag} U = 1$. In the ordered phase the sigma model is treated in  leading order perturbation theory, writing $U = e^{ig\varphi^a \tau^a / 2}$ and expanding the exponential.
In the disordered phase,
%\footnote{The disordered phase of the sigma model that is, not the $SU(N)$ theory.}
the energy\footnote{Here and throughout, extensive quantities are written per unit volume.}
is minimised close to the phase transition with $M \simeq M_c$,
\begin{gather} \label{disu}
U = - \frac{N^2 M_{c}^{4}}{30 \pi^2}
\end{gather}
and $g^2 (M_c) = \pi^2/N$.
In the ordered phase, one obtains
\begin{gather} \label{ordu}
U = \frac{N^2 M^4}{120 \pi^2}\, ,
\end{gather}
so that the energy is indeed minimised at $M \simeq M_c$, on the disordered side of the sigma model phase transition.

The extension to finite temperature was discussed in \cite{Kogan:2002yr}.
At finite temperatures, the energy minimisation argument is modified: one must consider the balance between energy $U$
and entropy $S$, minimising the free energy  $F= U-TS$. Since the parameter $H$ corresponds to thermal disordering, one expects
generically that $S$ will vanish for vanishing $H$. In the $SU(N)$ theory at moderate temperatures, the 
degrees of freedom correspond to glueballs. Since these are heavy, the excitations (disordering), and consequently the entropy, will be small.
One can thus attempt to calculate the entropy as some expansion in the small parameter $H$.

The leading order contribution in $H$ is \cite{Kogan:2002yr} a term of the form $H \log H$, multiplied by a
coefficient which is an $SU(N)_L \otimes SU(N)_R$ symmetric correlator of $U$ fields.
In the disordered (symmetric) phase of the sigma model, this expectation value vanishes.
Furthermore, since the leading order contribution in $H$ to the energy is positive definite, one finds that
the free energy is minimised with $H = 0$ at the minimum of the \emph{energy}.

Thus, in the disordered phase of the sigma model, the minimum of the free energy is at $M \simeq M_c$ with, from (\ref{disu}),
\begin{gather}
F = - \frac{N^2 M_{c}^{4}}{30 \pi^2}\, .
\end{gather}
In the ordered phase of the sigma model, the leading contribution to the entropy at small $H$ is
\begin{gather}
S = - \frac{N^2 M^3}{6 \pi^2} H \log H\, .
\end{gather}
Then, from (\ref{ordu}),
\begin{gather}
F = \frac{N^2 M^{4}}{120 \pi^2} + T \frac{N^2 M^3}{6 \pi^2} H \log H\, .
\end{gather}
Minimising with respect to $H$ and $M$, one finds that $F$
is minimised in the sigma model disordered phase (with $\langle U \rangle = 0$)
from $T=0$ up to $T = T_c \simeq 0.33 M_c$, beyond which $F$ is minimised with $M$ in the ordered phase 
of the sigma model (with $\langle U \rangle \neq 0$).
Since $U$ plays the same role as the Polyakov loop variable at finite temperature,
this corresponds to a deconfinement phase transition in the pure $SU(N)$ gauge theory.

As a result of the minimisation procedure one finds that the dimensionless quantity $H/M$ is equal to $e^{-1}$. This raises the question of whether neglected terms of $O(H)$, which have the same magnitude as the retained terms of $O(H \log H)$, could considerably affect the calculation. It would, therefore, be desirable to extend the calculation
to higher order in $H$. This we do in the remainder of this paper.
\section{Extended analysis} \label{sec:ext}
We wish to extend the previous calculation to include terms beyond the leading order in the kernel corresponding to thermal disorder $H$.
Although it is not at all clear a priori how one might do this in general, we shall, by providing an alternative procedure to that underlying the results of \cite{Kogan:2002yr,Kogan:1994wf} outlined above, show that an extended analysis is indeed possible within the context and inherent approximations of the Kogan--Kovner model. It should be recalled that the stumbling block to any improvement is the calculation of the entropy, $S = -\mathrm{tr} \rho \log \rho$.

Let us try the following gambit. Instead of restricting ab initio the density matrix to the form (\ref{qcdans}), 
imagine that we take some arbitrary gauge-invariant density matrix ansatz depending on the $A$ fields and integrated over the $U$ fields.
We allow this new ansatz (and whatever kernels it may contain) to remain arbitrary until we have no choice but to restrict it.
Now we integrate out the A fields to obtain a partition function of $U$ fields with respect to some action.

Next we introduce a separation of
momenta into high and low modes with $k \lessgtr M$
and integrate out the high mode $U$ fields as before. This effects a renormalisation group transformation on the low modes,
replacing the bare coupling $g^2$ --- which is not arbitrary, since it is defined by the gauge transformations (\ref{gt}) --- by the running coupling
$g^2(M)$. 
Now provided our ansatz is sufficiently close to the correct density matrix for $SU(N)$, the theory will be asymptotically free.
We are thus left with an action for the low
modes which is again some complicated sigma model, with a renormalised coupling $g^2(M)$ which we expect to be small provided $M$
is large and vice versa.

Now consider this model as a statistical mechanical model at `temperature' $g^2(M)$. We make the plausible assumption 
that this sigma model will, as $M$ is varied, undergo a symmetry-breaking transition
at `temperature' $g^2(M_c)$ from a `thermally disordered' (symmetric) phase at large $g^2(M_c)$ to an ordered phase at small $g^2(M_c)$.
Further, it is clear --- since the Polyakov loop $\langle U \rangle$ is zero in the former phase and non-zero in the latter ---
that this sigma model phase transition corresponds directly to the deconfinement transition in the $SU(N)$ theory. 

This argument is quite general; on review, it is clear that our only assumptions are that the ansatz is sufficiently close to $SU(N)$
and that the low mode sigma model undergoes a symmetry-breaking phase transition. In particular, let the ansatz, which is arbitrary and need not be Gaussian, be the \emph{correct} density matrix for $SU(N)$. The first assumption is certainly true. 
If the second assumption is also true, then we have constructed an exact argument that the deconfinement transition in $SU(N)$
corresponds to the phase transition in the low mode sigma model.

Thus, in order to study deconfinement in $SU(N)$, our aim should be to model the physics of each sigma model phase as 
accurately as possible and calculate the transition scale $M_c$. We then
calculate the free energy of $SU(N)$ in each phase, 
including any possible contribution from the high modes, at temperature $T$ and extract the minimal free energy. The deconfinement 
transition occurs
at the temperature for which the free energies calculated in the ordered and disordered phases of the low mode sigma model coincide.

Although we will take (\ref{qcdans}) as the ansatz for the density matrix, we shall keep the kernels $G^{-1}$ and $H$ arbitrary until we have no choice but to restrict them.

In the disordered phase no progress seems possible without restricting the arbitrary kernels. Following \cite{Kogan:2002yr},
we adopt the forms (\ref{rkers}) as before and the analysis is identical. 
The Boltzmann factor is $e^{-M_g/T}$ in this case where $M_g$ is the lightest glueball mass, so we expand the small entropy to leading order
and get zero as before. The resulting minimal free energy is thus independent of the temperature and we find
\begin{gather} \label{minlf}
F = - \frac{N^2 M_{c}^{4}}{30 \pi^2}\, ,
\end{gather}
where $M_c \simeq 1.33 \mathrm{GeV}$ is the sigma model transition scale predicted by the mean field calculation of \cite{Kogan:1994wf}. 

In the leading order perturbation theory approximation to the ordered phase of the  sigma model, however, minimisation with respect to arbitrary kernels $G^{-1}$ and $H$ for both high and low modes is possible. Further, the analysis can, as desired, be carried out to all orders in the thermal disorder kernel $H$.

In this approximation, the $U$ matrices can be parameterised in the standard exponential form and expanded in the coupling $g$
\begin{gather}
U = \exp\bigg\{ig \varphi^a \frac{\tau^a}{2}\bigg\} = 1 + ig \varphi^a \frac{\tau^a}{2} + \dots  
\end{gather}
Hence at leading order one can take
\begin{align}
U &\simeq 1\, , \nonumber \\
\partial_i U &\simeq  ig \partial_i \varphi^a \frac{\tau^a}{2}. 
\end{align}
Thus, the gauge transformations (\ref{gt}) reduce to
\begin{gather}
A^{a}_{i} \rightarrow A^{a}_{i} - \partial_i \varphi^a  
\end{gather}
and the Hamiltonian (\ref{ham}) reduces to 
\begin{gather}
\mathcal{H} = \frac{1}{2} \left[ E^{a2}_{i} + (\epsilon_{ijk} \partial_j A^{a}_{k})^2 \right]\, . 
\end{gather}
But these last two equations describe the theory $U(1)^{N^2-1}$: in the leading order of sigma model perturbation theory, the 
$SU(N)$ Yang--Mills theory reduces to the $U(1)^{N^2-1}$ free theory.
Moreover, the density matrix (\ref{qcdans}) becomes Gaussian again, because the gauge transformations are linear. One has
\begin{gather} \label{gans}
\rho [A,A^{'}] = \int D\varphi \; 
\exp\bigg\{-\frac{1}{2} \left[ A G^{-1} A + (A' - \partial \varphi) G^{-1} (A' - \partial \varphi) - 2 A H (A' - \partial \varphi) \right] \bigg\}\, . 
\end{gather}

Now the theory of $N^2-1$ $U(1)$ free fields in $3+1$ dimensions 
is completely tractable; the variational analysis for the $U(1)$ theory
(with Gaussian ansatz (\ref{gans})) was discussed in \cite{Gripaios:2002xb}. The free energy in momentum space in terms of the arbitrary kernels
$G^{-1}$ and $H$ is
\begin{multline}
F = \frac{N^2-1}{2} \int \frac{d^3p}{(2\pi)^3} 
\bigg[  G^{-1}(1 + GH) + p^2 G (1 - GH)^{-1} 
 \\
 - 4T \left(  \log \left[ \frac{GH}{  (1-(GH)^2)^{1/2} - (1-GH)}\right] 
- \log \left[ \frac{1- (1-(GH)^2)^{1/2}}{GH}\right] \left( \frac{1- (1-(GH)^2)^{1/2}}{(1-(GH)^2)^{1/2}-(1-GH)} \right) \right) \bigg]\, .
\end{multline}  
The kernels which minimise the free energy are
\begin{align} \label{mkers}
G^{-1} &= p 
\left( \frac{1+ e^{-\frac{2p}{T}}}{1 - e^{-\frac{2p}{T}}}\right), \nonumber \\
H &= 2p
 \left( \frac{e^{-\frac{p}{T}}}{1 - e^{-\frac{2p}{T}}}\right)
\end{align}
and the minimal value of the free energy at temperature $T$ is
\begin{align} \label{minpf}
F &= \frac{N^2-1}{\pi^2} \int_{0}^{\infty} p^2dp \; \left[ \frac{p}{2} + T \log (1 - e^{-p/T})\right] \nonumber \\
  &= - \frac{(N^2-1)T^4}{3 \pi^2} \int_{0}^{\infty} dx \frac{x^3}{e^{x} -1} \nonumber \\
  &= - \frac{\pi^2(N^2-1)T^4}{45}\, ,
\end{align}
where the zero-point term has been discarded.
All of this is of course consistent with the standard analysis of photon gases in statistical mechanics.

The minimal free energy of $SU(N)$ in the ordered phase of the sigma model at temperature $T$ is, from (\ref{minpf}) and dropping sub-leading contributions of $O(N^0)$,
\begin{gather}
F = - \frac{\pi^2 N^2 T^4}{45}.
\end{gather}
So we see that the free energy of $SU(N)$ is minimised with $M=M_c$ in the disordered phase of the sigma model for temperatures from zero up to
a temperature $T_c$ where
\begin{gather} 
F = - \frac{N^2 M_{c}^{4}}{30 \pi^2} =  - \frac{\pi^2 N^2 T_{c}^{4}}{45}\, ,
\end{gather}
which in turn implies 
\begin{gather}
T_c = \left( \frac{3}{2}\right)^{1/4} \frac{M_c}{\pi} \simeq 470 \mathrm{MeV}\, .
\end{gather}
We note that the transition temperature is shifted by only a very small amount compared to the result $T_c \simeq 450 \mathrm{MeV}$ obtained in \cite{Kogan:2002yr}.
The calculation is improved in the sense that, in the high temperature phase of $SU(N)$, which corresponds to 
the ordered phase of the sigma model, we have been able to extend the original analysis to include all orders of the thermal disorder kernel.
This is desirable because at high $T$ this kernel, which corresponds to the Boltzmann factor,  is of order unity.
The calculation is also improved in that the minimal kernels in the high $T$ phase, approximated as free gluons, are the exact ones.
If we had performed the calculation with the kernels (\ref{rkers}), we would not have been able to obtain the true minimum of the free energy.
As in \cite{Kogan:2002yr}, we find that the deconfinement phase transition is strongly first order with latent heat $\Delta E=\frac{4\pi^2 N^2}{45} T_{c}^{4}$.

Finally, it will also be of interest to calculate the ratio of the transition temperature to the lightest glueball mass in the model, which is $2M_c$ \cite{Gripaios:2002bu}. One obtains
\begin{gather} \label{ratio}
\frac{T_c}{2M_c} = \frac{1}{2\pi}(\frac{3}{2})^{1/4} \simeq 0.18.
\end{gather}
\section{Discussion}\label{sec:disc}
In this extended variational analysis, we have identified a phase transition (within the approximations made) at 470 MeV.
This seems rather high in comparison with numerical simulations performed on the lattice, which give around 280 MeV for $SU(3)$ \cite{Teper:1998kw}. However, the estimate obtained for the transition temperature is only expected to be approximate since it is sensitive to the value of the critical scale $M_c$, which is calculated in the mean field approximation of the sigma model.This sensitivity can be removed by computing the \emph{ratio} of the transition temperature to the lightest glueball mass in the model (\ref{ratio}). One then obtains a value of 0.18, which agrees with the lattice result for $SU(3)$ to two significant figures. 

There are other reasons why we expect the calculation to be only approximate. The most important point to be aware of is that in the original zero temperature analysis, the $SU(N)$ gauge theory was
hived into two parts (the high and low modes) for the purpose of tractability. The former corresponds approximately to the perturbative gauge theory, 
which is well understood (and which we have treated in the zeroth order)
and the latter to the low energy sector, which is less well understood and is treated in the mean field approximation.
In considering the theory at finite temperature, the phase transition corresponds to a jump between the two sectors. 
So in doing the analysis, we are really asking the question:
at what temperature does the free $U(1)^{N^2-1}$ gauge theory become thermodynamically more favourable than the 
low energy theory calculated in the mean field approximation?

Whilst this is a perfectly sensible question, to which we have obtained a sensible answer, one must ask whether this means anything
for the full $SU(N)$ gauge theory. One is interpolating between a low energy theory, which is already only approximate, and a high energy
theory which is only correct in the ultra-violet limit. This is, to say the least, rather crude. However,
we believe that the \emph{principle} of the method is rather powerful, in that there is scope to improve upon the calculation.
The simplest way in which this can be done is to include perturbative corrections to the free energy coming from the high modes.
The finite-temperature corrections should be added to (\ref{minpf}). In contrast, only the zero-temperature corrections should be added to (\ref{minlf}),
since there are no thermal contributions in this sector by assumption.

The second way in which improvements can be made is to improve corrections to the low mode sector. We believe that the crudest approximation here
is in taking the leading order of perturbation theory in the ordered phase of the sigma model. Clearly this is not appropriate close to the
sigma model phase transition, which corresponds also to the $SU(N)$ phase transition in this model. So if there are large shifts in the sigma model
behaviour near the transition, we would expect the transition temperature to be significantly shifted. 
An improved treatment of the sigma model near the phase transition necessarily calls for a higher order or non-perturbative calculation to be performed. But then one has to calculate the entropy for a non-free theory. Such a task is beyond our present calculational abilities. However, we are encouraged by the fact that the all orders in $H$ result for the free energy in the ordered phase obtained in this paper differs only very slightly near the phase transition from the one obtained in \cite{Kogan:2002yr} where only the leading $H \log H$ contribution was taken into account. This is, of course, why the transition temperature is not significantly shifted by the improved analysis. Now a non-perturbative calculation of the entropy to order $H \log H$ does seem to be possible, and is currently under way \cite{us}. 

\section{Acknowledgments}
We thank I. Kogan and A. Kovner for their contribution during the early stages of this work.

\end{document}